# A Background-Free Direction-Sensitive Neutron Detector

Alvaro Roccaro,[1*] H. Tomita,[1] S. Ahlen,[1] D. Avery,[1] A. Inglis,[1] J. Battat,[2] D. Dujmic,[2] P. Fisher,[2,4,5] S. Henderson,[2] A. Kaboth,[2] G. Kohse,[2] R. Lanza,[2] J. Monroe,[2] G. Sciolla,[2] N. Skvorodnev,[3] H. Wellenstein,[3] and R. Yamamoto[2]

[1]*Physics Department, Boston University, Boston, MA 02215,* [2]*Physics Department and Laboratory for Nuclear Science, Massachusetts Institute of Technology, Cambridge, MA 02139,* [3]*Physics Department, Brandeis University, Waltham, MA 02454,* [4]*MIT Kavli Institute and Institute for Soldier Nanotechnology, Massachusetts Institute of Technology, Cambridge MA 02139,* [5]*MIT Kavli Institute, Massachusetts Institute of Technology, Cambridge MA 02139* [*]*Deceased*

**We show data from a new type of detector that can be used to determine neutron flux, energy distribution, and direction of neutron motion for both fast and thermal neutrons. Many neutron detectors are plagued by large backgrounds from x-rays and gamma rays, and most current neutron detectors lack single-event energy sensitivity or any information on neutron directionality. Even the best detectors are limited by cosmic ray neutron backgrounds. All applications (neutron scattering and radiography, measurements of solar and cosmic ray neutron flux, measurements of neutron interaction cross sections, monitoring of neutrons at nuclear facilities, oil exploration, and searches for fissile weapons of mass destruction) will benefit from the improved neutron detection sensitivity and improved measurements of neutron properties made possible by this detector. The detector is free of backgrounds from x-rays, gamma rays, beta particles, relativistic singly charged particles and cosmic ray neutrons. It is sensitive to thermal neutrons, fission neutrons, and high energy neutrons, with detection features distinctive for each energy range. It is capable of determining the location**

**of a source of fission neutrons based on characteristics of elastic scattering of neutrons by helium nuclei. The detector we have constructed could identify one gram of reactor grade plutonium, one meter away, with less than one minute of observation time.**

We report a significant advance in the detection of neutrons and determination of their energy and direction of motion in a simple, compact device. The concept [1-3] is illustrated in Figure 1. The detector is in a chamber that contains $CF_4$ gas at low pressure. One bar or more of $^4He$ is added to provide a target for neutrons through elastic scattering, and a few torr of $^3He$ is added to detect thermal neutrons via the reaction $n+^3He \rightarrow ^3H+p$. The relative amounts reflect the 8b cross section for elastic scattering of neutrons on $^4He$ and the 5327 barn cross section for the absorption of thermal neutrons by $^3He$. A cathode mesh and field cage set up an electric field in an electron drift region, and a grounded mesh is separated by a few hundred microns from an anode copper plate that sets up a high electric field in an amplification region. A charged particle moving through the chamber leaves a trail of ionization electrons in its wake. The electrons from this track drift to the amplification plane, where an electron avalanche occurs, accompanied by the emission of scintillation light from $CF_4$. The scintillation light is imaged by the lens[1] and CCD camera[2]. If a neutron collides elastically with a $^4$He nucleus it produces a track similar to that of an alpha particle from a radioactive source, except at lower energy. If a neutron is absorbed by a $^3$He nucleus it produces back-to-back tracks of a proton and triton with 764 keV of kinetic energy. The device is blind to minimum ionizing charged particle tracks and has image formation and clearing time of the order of 10 ns. Determination of the third component of tracks can be made with the use of a photomultiplier tube to monitor drift time, which is up to 3 μs.

---

[1] We use a Nikon lens with a focal length of 55 mm and an f number of 1.2. For the alpha particle images of Figure 2, the lens was 39 cm from the amplification mesh. For the images of Figure 3, the lens was 28 cm from the amplification mesh.

[2] The Apogee U6 camera used a Kodak KAF-1001E CCD chip, which has a 1024 x 1024 array of 24 x 24 micron pixels, and has a read noise of 7 electrons. The quantum efficiency is 65% at 630 nm (the peak of the $CF_4$ emission spectrum[2]). We used 4 x 4 hardware binning for the pixels.

Our concept is similar to, but much simpler than, one proposed [4] in 1994 to search for dark matter. That device utilized a time projection chamber with parallel plate avalanche counter, an optical readout using the light-emitting gas triethylamine (with 280 nm wavelength), and a 4.5 kG magnetic field to reduce electron diffusion. The light collecting system included an ultraviolet lens and an image intensifier that was followed by a phosphor screen. The image on the phosphor screen was viewed by a second lens and a CCD video camera. Another device similar to ours has been built [5] to detect thermal neutrons using a Gas Electron Multiplier (GEM) in a 5 cm diameter chamber with a 2 cm drift distance. The gas used was a mixture of $CF_4$ (300 torr) and $^3$He (460 torr). Our early work showed that commercial 50 μm thick GEMs were too thin to use with low pressure $CF_4$, which is why we adopted the mesh approach. In addition, GEMs are quite expensive and fragile in comparison to the meshes that we use.

Other standard neutron detectors currently in use include the $^3$He based Bonner sphere technique [6], pulse shape discrimination in organic scintillators to distinguish elastic collisions of neutrons with protons from gamma ray signals [7-8], and various schemes making use of other exoergic thermal neutron capture reactions (including e.g.

$n + {}^6\text{Li} \rightarrow \alpha + {}^3\text{H} + 4.786 \text{ MeV}$, or $n + {}^{10}\text{B} \rightarrow \alpha + {}^7\text{Li} + 2.79 \text{ MeV}$). For fast neutrons, our imaging technique is superior because no moderator is needed and the image of the recoil track gives both energy and direction information about the incident neutron. For thermal neutrons, our imaging method gives the interaction location within the detector with 200 μ precision and is insensitive to gamma and beta rays, both of which are important for background rejection.

We recently exposed a prototype detector to several sources of radiation, including an $^{241}$Am alpha source, a $^{252}$Cf fission spectrum neutron source[3], a deuterium-tritium neutron generator[4], and cosmic ray neutrons. We used an Apogee U6 CCD camera and a Nikon 55 mm lens, which provided a field of view 10 cm on a side for the neutron exposures. The field cage drift was 10 cm high, and the meshes that bounded the drift region were made of stainless steel. Figure 2 shows a pair of tracks from the alpha source. The field of view was 14 cm for this setup. The light intensity is indicated by the color scale, and the Bragg peak of ionization is apparent. The gap in one of the tracks is due to a nylon wire that separates the mesh from the copper anode. The direction of motion of a stopping alpha particle can be determined by the reduction of ionization as the alpha particle approaches its end of range.

Figure 3 shows 15 successive recoil events from collisions with $^{252}$Cf neutrons. The amplification plane was parallel to the floor of the lab and the source was six meters away, about the same height from the floor as the detector. The neutrons came from the bottom of the images. This can be observed from the orientation of the recoils and the diminishing of light intensity toward the end of range. One of the tracks is going the wrong direction, and is most likely a neutron that scattered off the wall of the lab. The excellent directionality information is due to the fortuitous circumstance that the differential elastic cross section for n –$^4$He scattering in the MeV region is largest for center-of-mass scattering angles of 180 degrees. Each of the images of Figure 3 had a one-second exposure time, and about one out of every three exposures had a neutron recoil event. The rate of recoil events from cosmic rays was observed to be about one per hour prior to the introduction of the neutron source.

---

[3] The $^{252}$Cf source emitted 12 million neutrons/s. Approximate cosmic ray neutron rates[8] are 10 thermal neutrons/m²/s, 20 neutrons/m²/s for neutrons with energy in the MeV range, and 20 neutrons/m²/s for neutrons with energy > 20 MeV.

[4] The Thermo MF Physics A-325 neutron generator produced 14.1 MeV neutrons at the rate of 50 million neutrons/s.

We have developed image analysis software to automate data analysis. Tracks are found and measured and scatter plots of projected track length vs. total energy are obtained[5]. This allows different particle types to be identified. Improved resolution will be possible with the use of photomultiplier tube information, which enables the determination of total track length. Figure 4 shows scatter plots for three cases: (a) 80 torr $CF_4$ ($V_{amp}$ = 700 volts, $V_{drift}$ = 250 volts/cm), 17 hr exposure to cosmic rays; (b) 80 torr + 1 torr $^3$He, 18.5 hr exposure to $^{252}$Cf + 5 hr to cosmic rays ($V_{amp}$ = 750 volts, $V_{drift}$ = 250 volts/cm); (c) 80 torr $CF_4$ + 560 torr $^4$He, 18.5 hr exposure to $^{252}$Cf + 5 hr to cosmic rays ($V_{amp}$ = 750 volts, $V_{drift}$ = 250 volts/cm) and (d) 80 torr $CF_4$ + 560 torr $^4$He + 1 torr $^3$He, 18.5 hr exposure to $^{252}$Cf + 5 hr to cosmic rays ($V_{amp}$ = 750 volts, $V_{drift}$ =

---

[5] Track-finding for the image analysis includes background subtraction to eliminate hot pixels, flat field corrections, median filters, thresholding and the application of Hough transforms to search for lines in images. Once tracks are found and thier0 lengths are determined from the Hough transform information, their pixel-by-pixel light levels are determined, and the total light level, which is proportional to energy loss, is measured.

250 volts/cm). The energy scales were determined from calibrations with alpha source particle tracks.

In Figure 4d we see two bands, the upper being due to neutron $^4$He recoils, and the lower being due to the triton + proton tracks from the n + $^3$He capture reaction. These tracks are only partially contained in the volume of the detector so its energy is not fixed at 764 keV. The elastic cross section for n + $^4$He elastic scattering is 8 barns at 1 MeV, and the thermal neutron capture cross section on $^3$He is 5327 barns. Thus, the detection efficiencies are about the same for fission and thermal neutrons and the ambient thermal neutron flux in the lab is found to be about the same as the fission neutron flux. Figure 4a shows background events. These are of four main types. (i) The left side of the plot has a population of very short tracks with large energy which are due to background radiation interacting directly with the CCD. We have found by visual inspection that these can be eliminated by track morphology cuts. They may also be eliminated by using a photomultiplier tube or electrical readout of the amplification plane in coincidence with the CCD exposure. (ii) most of the plot is populated by a broad swath of events with ranges of a cm or more, and energies of 1/2 MeV or more. These are due to alpha particles from radioactive nuclei in the decay chains of U/Th contaminants found in chamber materials. Most of these are coming from the field cage and visual inspection indicates they may be removed by rejecting recoil events that have part of their track near the edge of the active region of the detector. Reduced backgrounds can be obtained with the use of copper for the field cage and mesh instead of stainless steel. (iii) there is a cluster of events with several mm of projected range and less than 100 keV of energy. These are recoil nuclei from elastic scattering of cosmic neutrons with $^4$He, as can be verified by examining Figure 4d which has no such feature. These events may be removed with a small loss in efficiency by a minimum projected range cut at 5 mm as may be seen by comparing Fig. 4a with Fig. 4c. (iv) the events at the bottom of Figure 4a are due to x-rays which are just barely detectable in pure $CF_4$ which has a gain about twice that of the helium mixtures. Owing to their low specific energy loss, these events are below threshold in a $CF_4$-helium mixture. We have developed algorithms to remove these backgrounds that will be reported in a forthcoming publication.

Figure 5 shows the distribution azimuthal angles of helium nuclei recoils from an exposure to the $^{252}$Cf source located at 90° with respect to the detector. The large peak at 90 degrees corresponds to recoils caused by neutrons coming from that direction. The peak near 270 degrees result from neutron that interact in the detector after recoiling from the wall immediately behind. Relative sizes of the two peaks agree with an estimated 13% of neutrons being reflected from the back wall. The smaller peaks in the picture are consistent with neutron reflections from the side walls of the shielded room.

We have built a large detector with a 20 cm drift field cage and a 30 cm diameter amplification plane (using copper mesh), which was viewed by a Schneider 17 mm focal length lens and an Andor iXon 888 EMCCD. This type of CCD has on-chip amplification which enables single photon per pixel imaging. Figure 6 shows two images from this device, with a gas mixture of 40 torr $CF_4$ + 600 torr $^4$He, 620 volts on the anode and a drift field of 125 V/cm. The one-second duration images were taken with the Cf source at six meters distance. Neutrons came from the top of the images. With a gas pressure of 4 bar, this same device would produce a similar image as these for a one-minute exposure time if one gram of reactor grade plutonium were one meter away. Figure 6 shows an image from the EMCCD setup for 14.1 MeV neutrons from the neutron generator. A number of inelastic interactions become possible at these higher energies, and identifying these interactions enables measurements of the high energy neutron environment. The central event shows three alpha particles emerging from the inelastic collision of a neutron with a carbon nucleus. The neutron energy threshold for this reaction is 7.9 MeV.

We are currently designing an EMCCD[6] based portable neutron detector with a detection volume of 20 liters that will operate with a mixture of $CF_4$, $^4He$, and $^3He$ at a pressure of 4 bars. The device, sketched in Figure 8, will be built and field tested for several of the applications mentioned above. The techniques we have described here are made possible by substantial improvements in CCD technology over the past decade. It is likely that further developments will continue to be made in the next several years that will lead to improved performance and to reduced costs for this new imaging technology. Benefits to many fields other than neutron detection will probably become possible.

*Much of this work was carried out by Alvaro Roccaro, who passed away unexpectedly during preparation of this letter. He will be sorely missed and we dedicate this letter to his memory.*

**References**

1. D. Dujmic et al., *Nuclear Instruments and Methods In Physics Research A* **584,** 327 (2008).

---

[6] The optics for the large detector include a Schneider 17 mm focal length lens with f number of 0.95 and an Andor iXon DU-888 EMCCD. The EMCCD camera utilizes an e2V CCD201-20 chip. The chip is back-illuminated with a quantum efficiency of 95% at 630 nm, and has a 1024 x 1024 array of 13 x 13 micron pixels. We used 2 x 2 hardware binning for the pixels.

**Acknowledgements** We acknowledge support of this work by grants from the National Science Foundation Department of Energy), Department of Homeland Security, the MIT Physics Department, the MIT Kavli Institute and the Institute for Soldier Nanotechnology at MIT.

**Author Information** Correspondence and requests for materials should be addressed to Peter Fisher (fisherp@mit.edu).

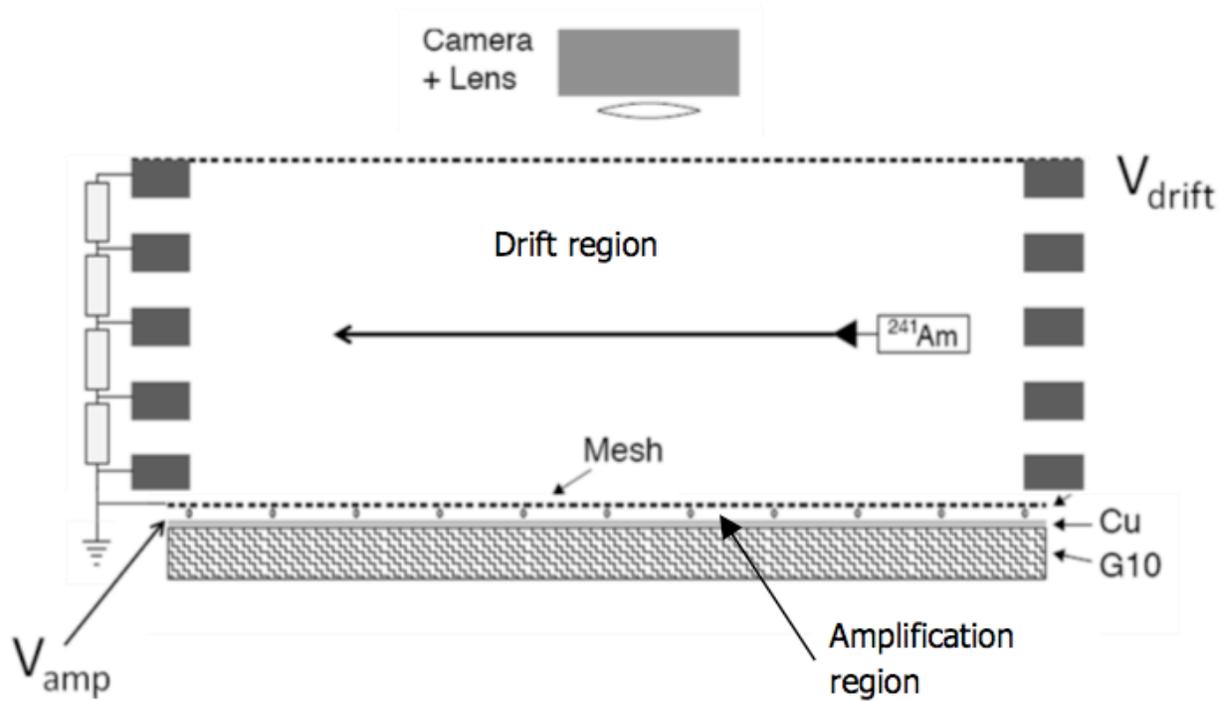

2 Figure 1. Schematic of neutron detector. $V_{amp}$ is applied to the amplification stage's
3 metal anode plate, and $V_{drift}$ is applied to the mesh at the top of the field cage that
4 sets up a uniform electric field in the electron drift region. An alpha particle source
5 is used to calibrate the device.

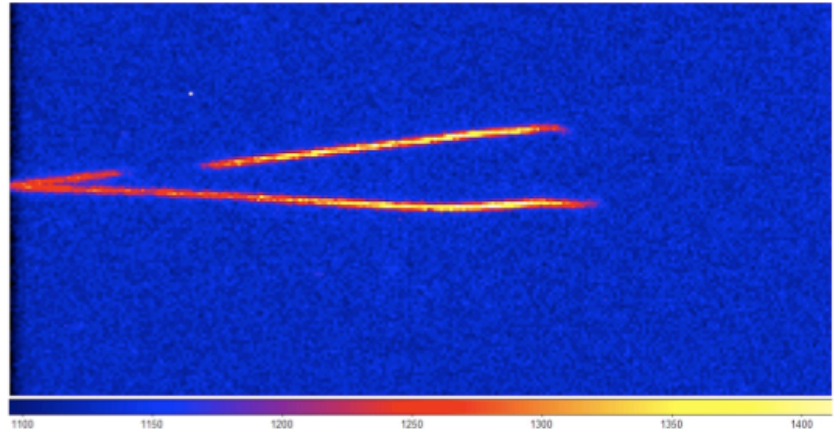

3   Figure 2. Two alpha tracks from $^{241}$Am source. Chamber gas was 80 torr $CF_4$, $V_{amp}$
4       = 740 volts, and $V_{drift}$ = −2500 volts. The width of the image is 14 cm.

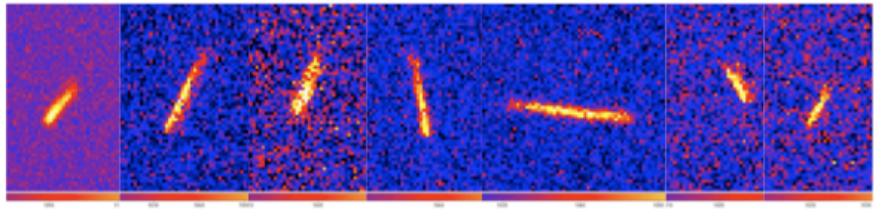
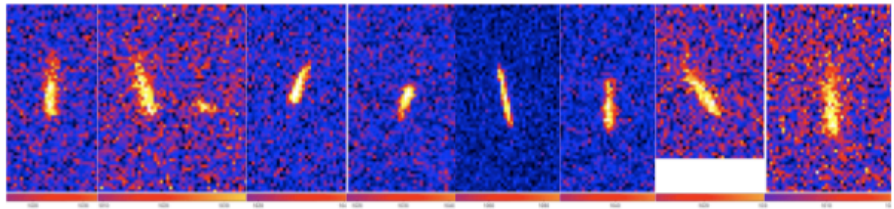

Figure 3. Fifteen successive recoil $^4$He tracks from $^{252}$Cf neutrons incident from the bottom of the image. Chamber gas was 80 torr $CF_4$ + 560 Torr $^4$He, $V_{amp}$ = 750 volts, and $V_{drift}$ = –2500 volts. The image heights are about 2 cm at the amplification plane.

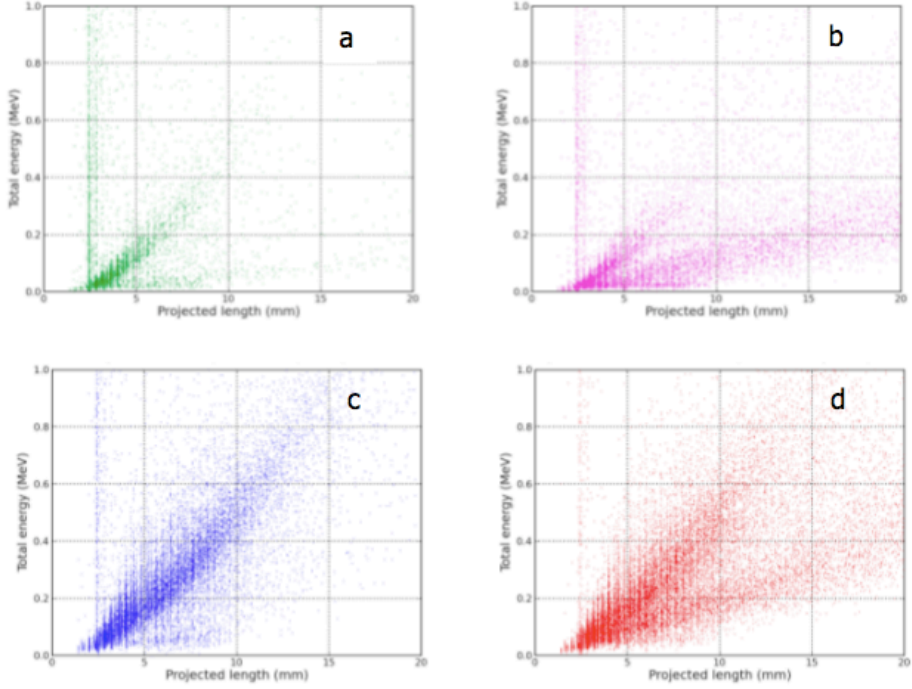

Figure 4. Total energy vs projected range for events for various exposure conditions. See text for discussion.

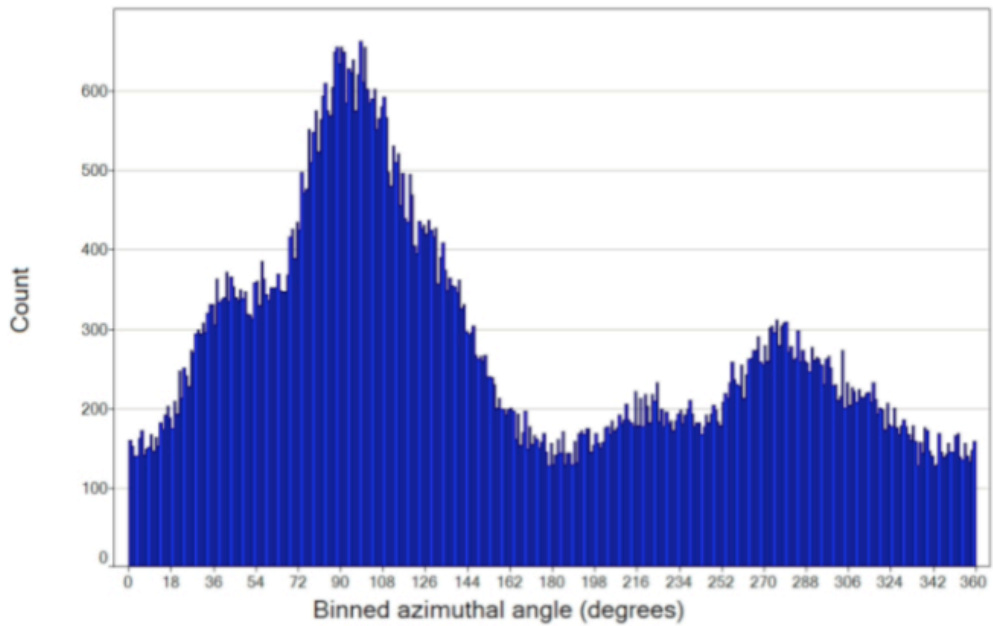

Figure 5 – Angular distribution of helium recoils from an exposure to at $^{252}$Cf neutron source located three meters from the detector. 90 degrees corresponds to nuclei recoiling away from the source.

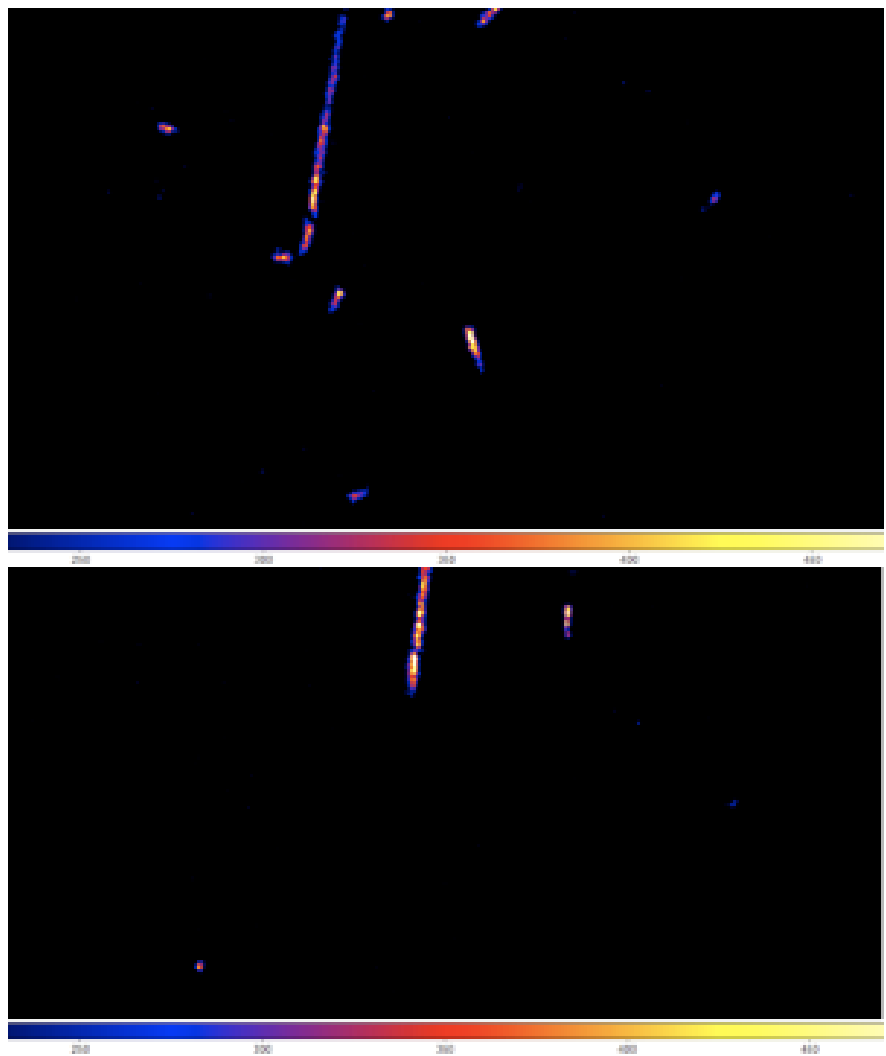

Figure 6. Two one second images of $^{252}$Cf neutron events for 30 cm diameter amplification plane with 20 cm drift distance, using 17 mm lens and EMCCD. The image size corresponds to the full 30 cm width of the amplification plane.

We are currently designing an EMCCD based portable neutron detector with a detection volume of 20 liters that will operate with a mixture of $CF_4$, $^4$He, and $^3$He at a pressure of 4 bars. The device, sketched in Figure 8, will be built and field tested for several of the applications mentioned above. The techniques we have described here are made possible by substantial improvements in CCD technology over the past decade. It is likely that further developments will continue to be made in the next several years that will lead to improved performance and to reduced costs for this new imaging

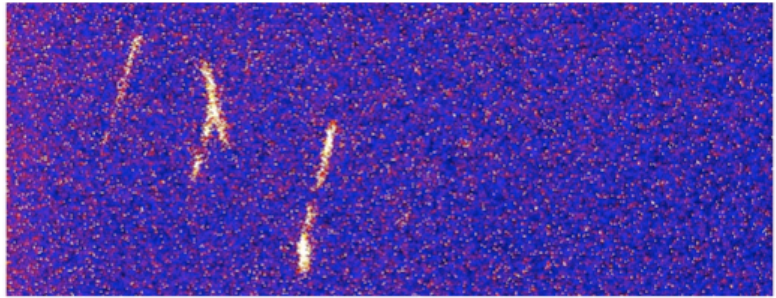

Figure 7. Image of three events with EMCCD for 14.1 MeV neutrons from neutron generator. The central event is probably due to the inelastic reaction $^{12}C(n, n'+3\alpha)$ which has an energy threshold of 7.9 MeV.

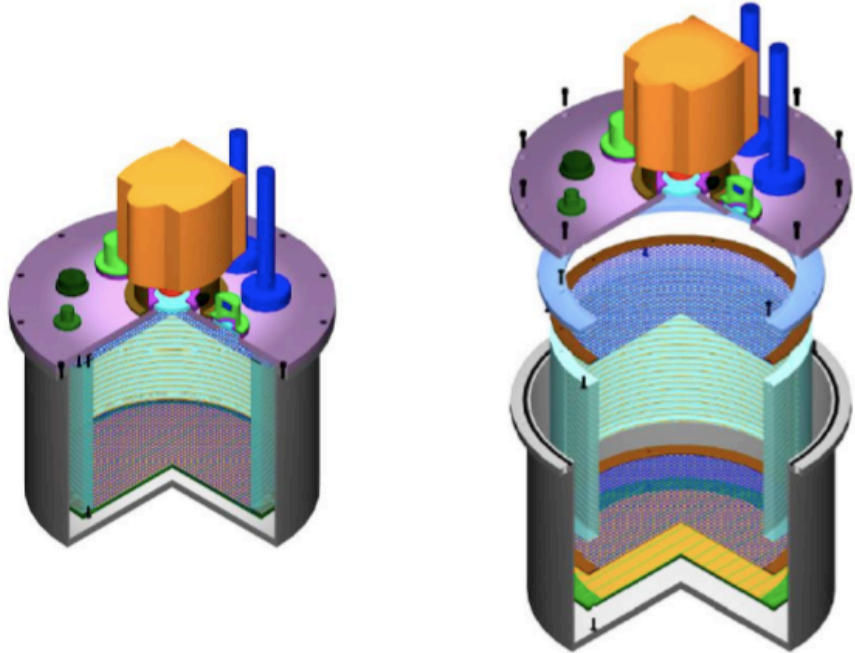

Figure 8. Sketch of 20 liter neutron detector. Exploded view is on right. The pressure vessel contains the field cage, the upper cathode mesh, the lower grounded amplification mesh and the anode plate. The Andor EMCCD with the Schneider 17 mm f/0.95 view the amplification plane through a window. High voltage feed-through connectors, a photomultiplier tube, and gas fittings are also shown in the top plate.